\documentclass[twocolumn,prb,aps,notitlepage,nofootinbib]{revtex4-2}

\usepackage[utf8]{inputenc}
\usepackage{array}
\usepackage[dvipsnames]{xcolor}
\usepackage[euler]{textgreek}
\usepackage{amsmath}
\usepackage{amsthm}
\usepackage{amsfonts}
\usepackage{amssymb}
\usepackage[a4paper,width=170mm,top=25mm,bottom=20mm]{geometry}
\usepackage{graphicx}
\graphicspath{ {Figures/} }
\usepackage{float}
\usepackage[export]{adjustbox}
\usepackage{hyperref}
\hypersetup{colorlinks = true, citecolor = blue, linkcolor= blue}
\usepackage{enumitem}
\usepackage{slashed}
\usepackage[customcolors,shade]{hf-tikz}
\usepackage{bm}
\usepackage{bbm}
\usepackage[title]{appendix}
\usepackage{mathrsfs}
\usepackage{tikz}
\usepackage{footmisc}
\usepackage{verbatim}
\usepackage{upgreek}
\usepackage{dsfont}
\usepackage{bbold}
\usepackage[section]{placeins}
\usepackage{natbib,hyperref}
\usepackage[normalem]{ulem}
\usepackage{lineno}

\usepackage{cleveref}

\usepackage{physics}
\usepackage{mathtools}
\usepackage{setspace}
\usepackage[makeroom]{cancel}

\newcommand{\ti}[1]{\textit{#1}}

\newcommand{\be}{\begin{equation}}
\newcommand{\ee}{\end{equation}}
\newcommand{\bes}{\begin{equation*}}
\newcommand{\ees}{\end{equation*}}
\newcommand{\f}[2]{\frac{#1}{#2}}

\makeatletter
\newcommand{\vast}{\bBigg@{3}}
\newcommand{\Vast}{\bBigg@{5}}
\makeatother

\theoremstyle{definition}

\theoremstyle{plain}

\begin{document}

\title{Nonlinear Hall effect in Weyl semimetals induced by chiral anomaly}
\author{Rui-Hao Li$^{1}$}\email{ruihao.li@case.edu}
\author{Olle G. Heinonen$^{2,3}$}
\author{Anton A. Burkov$^{4}$}
\author{Steven S.-L. Zhang$^{1}$}\email{shulei.zhang@case.edu}
\affiliation{$^1$Department of Physics, Case Western Reserve University, Cleveland, Ohio 44106, USA\\
$^2$Materials Science Division, Argonne National Laboratory, Lemont,
Illinois 60439, USA\\
$^3$Northwestern-Argonne Institute of Science and Engineering, Evanston,
Illinois 60208, USA\\$^4$Department of Physics and Astronomy, University of Waterloo, Waterloo,
Ontario N2L 3G1, Canada
}

\begin{abstract}
We predict a nonlinear Hall effect in certain Weyl semimetals with broken inversion symmetry. When the energy dispersions about pairs of Weyl nodes are skewed---the Weyl cones are ``tilted"---the concerted actions of the anomalous velocity and the chiral anomaly give rise to the nonlinear Hall effect. This Hall conductivity is linear in both electric and magnetic fields, and depends critically on the tilting of the Weyl cones. We also show that this effect does not rely on a finite Berry curvature dipole, in contrast with the intrinsic quantum nonlinear Hall effect that was recently observed in type-II Weyl semimetals.

\end{abstract}

\date{\today}
\maketitle

\maketitle

\section{Introduction}

Weyl semimetals (WSMs) \cite{Ashvin11PRB_TSM-Fermi-arc, Burkov&Balent11PRL_WSM, yRan11PRB_Hall-WSM, gXu11PRL_HgCrSe-magnWSM, Halasz12_TRI_WSM, Huang15NatCom_TaAs, bqLv5PRX_WSM-TaAs, Lv2015NatPhys_TaAs, hmWeng15PRX_WSM-noncen, Hasan15Science_FermiArc} are a newly discovered class of quantum materials which can host a number of exotic massless quasiparticles called Weyl fermions with a well-defined chirality near the band-crossing points (Weyl nodes). One of the most unique features of Weyl fermions is the chiral anomaly \cite{Adler69PR_chiral-anomaly, Bell&Jackiw69_Chiral-anomaly} – breaking of the chiral symmetry at the quantum level leading to the nonconservation of chiral charges. The manifestation in WSMs is that a pair of Weyl nodes of opposite chiralities acts as source and drain of electrons in the presence of nonperpendicular electric and magnetic fields, resulting in a density difference between the two nodes, while preserving the total electron density \cite{Aji12_PRB_chiral_anomaly, son13PRB_NMR-WSM}.

To date, the most remarkable phenomenon induced by the chiral anomaly is the negative longitudinal magnetoresistance \cite{son13PRB_NMR-WSM, Kim14PRB_WSM_NMR, Burkov14PRL_WSM_NMR}, which was observed experimentally in WSMs such as TaAs \cite{xHuang15PRX_NMR-WSM, Zhang16NatComm_NMR_TaAs}. Intuitively, this phenomenon can be understood via the chiral magnetic effect~\cite{Fukushima08PRD_CME, Zyuzin12PRB_WSM_CME} in WSMs: In the absence of an electric field, there are equal numbers of Weyl fermions with opposite chiralities moving in opposite directions (collinear with the external magnetic field), which results in zero net charge current; when an electric field is applied along the magnetic field direction, an effective chemical-potential difference between Weyl fermions with opposite chiralities is created due to the chiral anomaly~\cite{son13PRB_NMR-WSM}, giving rise to an imbalance between the two fluxes of Weyl fermions and consequently a net charge current $\vb j\propto (\vb E\vdot\vb B)\vb B$. 

More recently, another related transport phenomenon induced by the chiral anomaly in WSMs called the planar Hall effect was proposed \cite{aBurkov17PRB_PHE-WSM, sTewari17PRL_PHE-WSM, Ma19PRB_PHE_tiltedWSM} and experimentally detected \cite{Kumar18PRB_PHE_GdPtBi, Chen18PRB_PHE_MoTe2, Liang19AIPAdv_PHE_MoTe2, Li19PRB_PHE_WTe2} , wherein the Hall current, the electric and magnetic fields are all coplanar. It is worth noting that both the negative magnetoresistance and the planar Hall effect in WSMs are linear responses to the external electric field.

\begin{figure*}[t]
	\includegraphics[width=\linewidth]{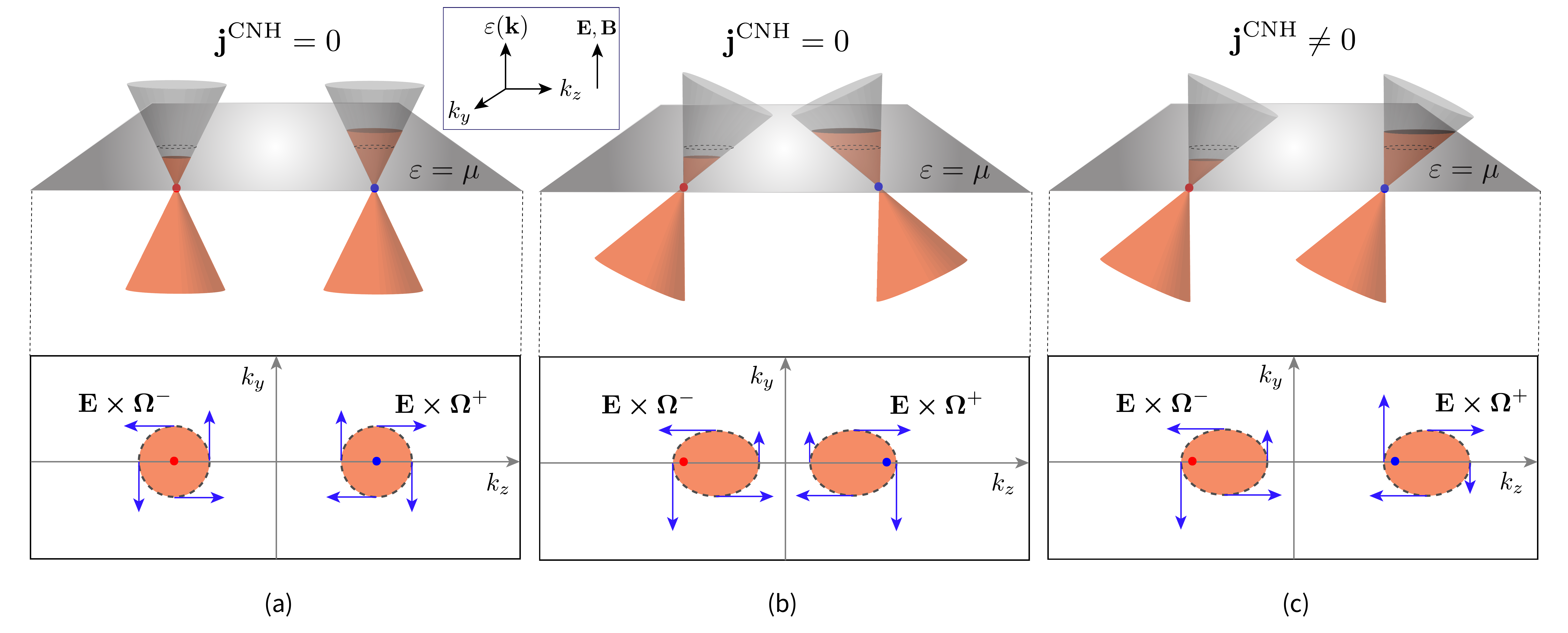}
    \caption{Schematics illustrating the physical mechanism of the chiral-anomaly-induced nonlinear Hall (CNH) effect for a pair of Weyl nodes of opposite chiralities, denoted by the blue ($s = 1$) and red ($s = -1$) dots. In the linear model used in our analysis, for a fixed value of $k_x$, each Weyl node has a linear dispersion along the $k_y$ and $k_z$ axes, forming a Weyl cone, as shown in the upper panel in each subfigure. For simplicity, the two Weyl nodes are separated along the $z$ axis and the external electric and magnetic fields are assumed to be in the $x$ direction. We show the scenarios where the pair of Weyl cones are (a) untilted, (b) tilted in opposite directions, and (c) tilted in the same direction. The gray horizontal planes cut through the energy dispersions at the equilibrium chemical potential $\mu$ with the corresponding Fermi-surface cross sections shown in the lower panels. The nonlinear Hall current arises as a consequence of the chiral anomaly and the anomalous velocity. On one hand, the chiral anomaly effectively leads to unequal electron densities between the two Weyl cones, as shown by the orange-filled parts of the cones. On the other hand, the anomalous velocity, whose direction and magnitude depend on the chirality of the Weyl cone as well as the location on the Fermi surface, is indicated by the blue arrows in the lower panels. In scenarios (a) and (b), the whole Fermi surface is symmetric about $\vb k = 0$, leading to a vanishing CNH current, whereas in scenario (c), an asymmetric Fermi surface leads to a finite CNH current.}
    \label{schematic}
  \end{figure*}

In this work, we predict another transport signature of the chiral anomaly in WSMs -- a nonlinear Hall effect with the Hall conductivity proportional to $\vb E\vdot\vb B$. 
The physical mechanism of the chiral-anomaly-induced nonlinear Hall (CNH) effect is illustrated in Fig.~\ref{schematic}, which shows a combined effect of the anomalous velocity and the chiral anomaly. It is well established that, in the presence of nonperpendicular electric and magnetic fields, the chiral anomaly results in a chiral-dependent modification of the electron density in the vicinity of each Weyl node~\cite{son13PRB_NMR-WSM}, i.e., $\delta n^s_{\vb{k}}\sim s\vb{E}\vdot\vb{B}$ with $s=\pm 1$ denoting the chirality. Moreover, due to the finite Berry curvature $\boldsymbol\Omega^s_{\vb{k}}$ of the Bloch states, the conduction electrons on the Fermi surface acquire an additional anomalous velocity $\vb v^s_a = \f{e}{\hbar}\vb{E}\cp\boldsymbol{\Omega}_{\vb k}^s$~\cite{Karplus54,Xiao2010}, which is perpendicular to the applied electric field. The direction of the anomalous velocity depends also on the chirality of the Weyl nodes. These two effects conspire to produce a nonlinear Hall current density $\vb{j}^\text{CNH}=-e\sum_{\vb{k},s}\delta n_{\vb{k}}^s \vb{v}_a^s$. 

As will be shown explicitly in Sec.~\ref{sec3}, a nonvanishing $\vb j^\text{CNH}$ requires WSMs with broken inversion symmetry. In addition, asymmetric Fermi surfaces are also necessary. One way to achieve this is via tilting of the Weyl cones, as demonstrated in Fig.~\ref{schematic}. It suffices to consider the simplest Weyl node configuration, that is, a pair of Weyl nodes with linear dispersions situated at the same energy level, to highlight the essential physics behind the CNH effect. In Fig.~\hyperref[schematic]{1(a)}--\hyperref[schematic]{1(c)} we consider three special cases for the pair of Weyl cones, which are untilted, tilted in opposite directions, and tilted in the same direction, respectively. As shown in the lower panels of Fig.~\ref{schematic}, which feature the projection of the Fermi surfaces onto the $k_x$-$k_y$ plane, in the first two cases, the whole Fermi surface for a pair of Weyl nodes is symmetric about $\vb k = 0$, resulting in a vanishing $\vb j^\text{CNH}$, whereas in the third case, the asymmetric Fermi surface leads to a finite $\vb j^\text{CNH}$ according to our calculation. 

The paper is organized as follows. In Sec.~\ref{sec2}, we set up the general formalism for evaluating the CNH current density in tilted WSMs, which involves a low-energy two-band Hamiltonian and the semiclassical Boltzmann equations. In Sec.~\ref{sec3A}, we present the general form of the CNH current density and its corresponding nonlinear response functions for both type-I and type-II WSMs. We explore multiple aspects of the CNH effect based on the features of the nonlinear response function and symmetry considerations. In Secs.~\ref{sec3B} and \ref{sec3C}, we discuss in detail the dependence of the CNH response on tilting and the relative energy shift of a pair of Weyl nodes, respectively.  In Sec.~\ref{sec3D}, we discuss the difference of the CNH effect from other nonlinear Hall effects that exist in the literature. In particular, we highlight the difference from the nonlinear Hall effect that arises from the Berry curvature dipole, which exists in time-reversal-invariant systems. Materials and experimental considerations for detecting the CNH effect will be discussed in Sec.~\ref{sec3E}. Lastly, we draw conclusions in Sec.~\ref{sec4}.

\section{\label{sec2}Formulation}

The Nielsen-Ninomiya fermion doubling theorem \cite{NN1981_NPB,NN1981_PLB} asserts that Weyl nodes must appear in pairs of opposite chiralities to ensure zero total chirality; thus, a WSM with broken inversion symmetry must have multiples of four Weyl nodes. As long as pairs of Weyl nodes are well separated in momentum space, we may examine the CNH effect by considering each pair independently. Without loss of generality, we assume that the Weyl cones are tilted along the $z$ axis. The simplest low-energy Hamiltonian for each Weyl node---a building block for realistic WSMs---is 
\be \label{eq1}
H^s_{\vb k} = \hbar v_F\left(s \vb k\vdot \boldsymbol\sigma+ R_s k_z\sigma_0 \right)+\mu_s,
\ee
where $\sigma_0$ is the $2 \times 2$ identity matrix, $\boldsymbol\sigma = (\sigma_x, \sigma_y, \sigma_z)$ are the three Pauli matrices, $v_F$ is the Fermi velocity, $s=\pm 1$ specifies the chirality of the Weyl node, the parameter $R_s$ characterizes the tilting of the Weyl cone, and $\mu_s$ denotes the energy shift of the Weyl node with chirality $s$. The Hamiltonian above can be obtained by linearizing a four-band model Hamiltonian for WSMs with broken inversion symmetry, as shown explicitly in Appendix~\ref{appx-A}.

For small tilting $\abs{R_s}<1$, the Fermi surface encloses only electron pockets (assuming that the chemical potential lies in the conduction band). In this case, the Hamiltonian describes a type-I Weyl node. When $\abs{R_s}>1$, which corresponds to a type-II Weyl node, unbounded electron and hole pockets are present at the Fermi energy~\cite{Soluyanov15Nat_typeIIWSM}.  It is also worth pointing out that alternatively, one may start with a low-energy Hamiltonian similar to \eqref{eq1} but takes into account the separation of the two Weyl nodes in momentum space $2Q$ on the $k_z$ axis. As presented in Appendix~\ref{appx-B}, this will lead to a $Q$-dependent nonlinear response. However, calculations show that the $Q$ dependence of the response function for type-II WSMs is generally very weak, and hence we may adopt the simpler Hamiltonian given by Eq.~\eqref{eq1}.

As mentioned previously, the presence of a finite Berry curvature gives a Weyl fermion an additional anomalous velocity. The Berry curvature is given by
\be \label{eq2new}
\boldsymbol{\Omega}^s_{\vb k} = i\bra{\grad_{\vb k}u_{\vb k}^s}\cp \ket{\grad_{\vb k}u_{\vb k}^s},
\ee
where $H^s_{\vb k}\ket{u_{\vb k}^s} = \varepsilon^s_{\vb k}\ket{u_{\vb k}^s}$. Furthermore, in the presence of a magnetic field, another physical quantity that can affect the electron wave-packet dynamics is the orbital magnetic moment given by \cite{Xiao2010} 
\be
\vb m^s_{\vb k} = -i\f{e}{2\hbar} \bra{\grad_{\vb k}u_{\vb k}^s}\cp \qty[H^s_{\vb k}-\varepsilon^s_{\vb k}]\ket{\grad_{\vb k}u_{\vb k}^s},
\ee
which arises from a self-rotation of the wave packet around its center of mass and modifies the energy dispersion as $\tilde \varepsilon^s_{\vb k} =\varepsilon^s_{\vb k}-\vb m_{\vb k}^s\vdot \vb B$.

With Hamiltonian~\eqref{eq1}, the energy dispersion of the Weyl node of chirality $s$ can be written as
\be \label{eq2}
\varepsilon^{s}_{\vb k} = \hbar v_F \qty(R_s k_z \pm k)+\mu_s,
\ee
where $+$ ($-$) sign corresponds to the conduction (valence) band and $k=\abs{\vb k}$. The Berry curvature and the orbital magnetic moment are then given by
\begin{subequations}
\begin{align}
	\boldsymbol{\Omega}^{s}_{\vb k} &= -s\frac{\pm\vb k}{2k^3}, \label{eq3} \\
	\vb m^s_{\vb k} &= -sev_F\f{\pm \vb k}{2k^2}. \label{eq5b}
\end{align}
\end{subequations}
The semiclassical equations of motion for a Weyl fermion wavepacket can be written as~\cite{Xiao2010}:
\begin{subequations}
\begin{align}
	D^{s}\dot{\vb{r}}^{s} &= \tilde{\vb v}^{s}_{\vb k}+\f{e}{\hbar}\vb E\cp\boldsymbol{\Omega}^{s}_{\vb k}+\f{e}{\hbar}(\tilde{\vb v}^{s}_{\vb k}\vdot\boldsymbol{\Omega}^{s}_{\vb k})\vb B, \label{eq6a}\\
	D^{s}\dot{\vb k}^{s} &= -\f{e}{\hbar}\vb E-\f{e}{\hbar}\tilde{\vb v}^{s}_{\vb k}\cp\vb B-\f{e^2}{\hbar^2}(\vb E\vdot\vb B)\boldsymbol{\Omega}^{s}_{\vb k}, \label{eq4b}
\end{align}
\end{subequations}
where $\tilde{\vb v}^s_{\vb k}\equiv \vb v^s_{\vb k}-(1/\hbar)\grad_{\vb k}(\vb m^s_{\vb k}\vdot\vb B)$, with $\vb v^{s}_{\vb k} = (1/\hbar)\grad_{\vb k}\varepsilon^s_{\vb k}$ being the group velocity of the Weyl fermion with chirality $s$, and $D^{s}\equiv 1+\f{e}{\hbar}(\vb B\vdot\boldsymbol{\Omega}^{s}_{\vb k})$ is the corresponding modified density of states. 

To compute the current density, we substitute Eq.~\eqref{eq4b} into the homogeneous steady-state Boltzmann equation with the relaxation-time approximation
\be \label{eq7new}
\dot{\vb k}^{s}\vdot \grad_{\vb k} f^s = -\f{f^{s}-f^{s}_{0}}{\tau},
\ee
and solve for $f^{s}\simeq f^{s}_{0}+f^{s}_{1}+f^{s}_{2}$, where $f^{s}_{0}$ is the equilibrium Fermi-Dirac distribution [at zero temperature $f^{s}_{0}=\Theta\left(\mu-\varepsilon^s_{\vb k}+\vb m^s_{\vb k}\vdot\vb B\right)$ with $\mu$ being the chemical potential], $f_1^s$ and $f_2^s$ are the corrections to the equilibrium distribution at the first- and second-order in electric field, respectively, and $\tau$ is the intranode relaxation time. We have also assumed that the internode scattering rate is much smaller than the intranode scattering rate $1/\tau$ and it can be neglected (see Appendix~\ref{appx-C} for proof). The current density can then be calculated via
\be \label{eq6}
\vb j = (-e)\sum_s \int_{\vb k}D^{s}\;\dot{\vb r}^{s}\:f^{s},
\ee
where $\int_{\vb k}$ is the shorthand notation for $\int d\vb k/(2\pi)^3$, the physical velocity $\dot{\vb{r}}^s$ is given by Eq.~\eqref{eq6a}, and $\sum_s$ represents summing the contributions from both Weyl nodes of opposite chiralities. 

Note that for a type-I WSM, we only need to calculate the contribution from the conduction (valence) band when the chemical potential lies above (below) the Weyl nodes, whereas for a type-II WSM, we need to sum the contributions from both the conduction and valence bands due to the emergence of the electron and hole pockets at the Fermi level. Moreover, due to the unbounded nature of the Fermi surface in a linear model, one needs to introduce a ultraviolet momentum cutoff $\Lambda$~\cite{Carbotte16PRB_tilt-WSM, Agarwal19PRB_tiltWSM} beyond which the linear model \eqref{eq1} can no longer be taken as a valid description of the WSM. Details of the calculation of nonlinear responses can be found in Appendix~\ref{appx-D}.

\section{\label{sec3}{Result and discussion}}
We find, up to $O(E^2B^1)$, that the CNH current stems from terms involving the Berry curvature $\boldsymbol{\Omega}^s_{\vb k}$ and is given by the following integral:
\be \label{eq7}
\vb j^\text{CNH} = \f{e^4\tau}{\hbar^2} \sum_s\int_{\vb k}\pdv{f_0^s}{\varepsilon^s_{\vb k}}\vb E\cp\boldsymbol{\Omega}^s_{\vb k}\qty(\vb E\cp\boldsymbol{\Omega}^s_{\vb k})\vdot(\vb v^s_{\vb k}\cp \vb B).
\ee
Note that the orbital magnetic moment $\vb m_{\vb k}^s$ is not responsible for the chiral-charge imbalance between a pair of Weyl nodes and hence does not contribute to $\vb j^\text{CNH}$.

By inspecting the structure of the integral above, it is evident that if the energy dispersion of the WSM is invariant under $\vb k\to -\vb k$, corresponding to a Fermi surface symmetric about $\vb k = 0$, the group velocity $\vb v^s_{\vb k} = (1/\hbar)\grad_{\vb k}\varepsilon_{\vb k}^s$ is an odd function of $\vb k$ and hence the integral over the reciprocal space vanishes. Thus, to obtain a nonzero $\vb j^\text{CNH}$, an asymmetric Fermi surface is necessary and in the current setup, it is provided by tilting of the Weyl cones as well as an unequal shift in energy between the two Weyl nodes, as shown in Eq.~\eqref{eq2}. In the following subsections, we discuss the CNH current density for tilted WSMs in details. 

\subsection{\label{sec3A} CNH response function}

Placing Eqs.~(\ref{eq2}) and (\ref{eq3}) in  Eq.~(\ref{eq7}) and evaluating the integral therein, the CNH current density for a tilted WSM can be expressed in the following form:
\be
\vb j^\text{CNH} = \sum_s \kappa^s (\vb E\vdot\vb B)(\vb E\cross\vu t),
\label{eq:J^CNH}
\ee
where $\vu t$ is the unit vector in an arbitrary tilt direction ($\vu t=\vu e_z$ in the present setup) and $\kappa^s$ is the nonlinear current response function for a Weyl node of chirality $s$. We find, for type-I WSMs
\be \label{eq10}
\kappa_\text{I}^s = \f{3e^4v_F^2\tau(\mu-\mu_s)}{40\pi^2\hbar\abs{\mu-\mu_s}^3}R_s,
\ee
and for type-II WSMs,
\be
\label{eq14}
\kappa_\text{II}^s = \f{5\kappa_\text{I}^s}{12 \abs{R_s}^5}\left(R_s^6+\f{3}{2}R_s^4-\f{1}{10}
 +\f{1-R_s^2}{\tilde\Lambda^2}+\f{3-2R_s^2}{2\tilde\Lambda^4}\right),
\ee
where we have introduced a dimensionless ultroviolet cutoff $\tilde\Lambda \equiv \hbar v_F\Lambda/(\mu-\mu_s)$ to deal with the open Fermi surface in the two-band model of type-II WSMs described by Eq.~\eqref{eq1} \cite{Carbotte16PRB_tilt-WSM, Agarwal19PRB_tiltWSM}. In real materials, the cutoff may be considered as an upper-bound of $\abs{\vb k}$ beyond which the bands cease to be linearly dispersing. It is worthy to point out that in type-II WSMs, when the energy corresponding to the momentum cutoff $\Lambda$ is much larger than the energy of the Weyl nodes relative to the Fermi energy, that is, $\tilde\Lambda \gg 1$, the cutoff-dependent terms in the response function $\kappa_\text{II}^s$ become negligible. This regime is desirable as we are mostly interested in the physics near the Weyl nodes, that is, when the Fermi energy is close to the Weyl-node energy. It further removes the dependence on the seemingly artificial cutoff $\Lambda$, making the result more universal.

 A few general remarks on the CNH effect are in order. First, the nonlinear Hall effect vanishes if the system is inversion-symmetric. This can be seen from the general form of the nonlinear Hall current $\vb{j}^\text{CNH}$ as given by Eq.~(\ref{eq:J^CNH}); the whole set of the external fields, i.e., $\left(\vb{E}\vdot\vb{B}\right)\vb{E}$,  is even under space inversion whereas $\vb{j}^\text{CNH}$ is parity-odd, so the response function must be zero if the system is invariant under space inversion. 
 
 Second, an untilted Weyl cone does not contribute to the CNH effect in the low-energy limit, which is indicated by Eqs.~\eqref{eq10} and \eqref{eq14}. Physically, this can be understood as follows: For an untilted Weyl cone, at any two $\mathbf{k}$ points on the pocket of the Fermi surface symmetric about the Weyl node, the anomalous velocity vectors have the same magnitude but point in opposite directions. This pairwise cancellation leads to a vanishing CNH contribution from the Weyl cone after summing over all states on the corresponding electron pocket, as shown schematically in Fig.~\hyperref[schematic]{1(a)}. More generally, a nonvanishing total $\vb{j}^\text{CNH}$ requires the whole Fermi surface to be asymmetric about the $\Gamma$ point, as we addressed earlier based on Eq.~\eqref{eq7}. This point is exemplified in Appendix~\ref{appx-A} where we explicitly calculate the total CNH response $\kappa$ for a four-band Weyl Hamiltonian with two pairs of Weyl nodes. 

 Third, as seen in Eqs.~\eqref{eq10} and \eqref{eq14}, the nonlinear Hall response $\kappa^s$ and hence the corresponding nonlinear Hall conductivity are proportional to $\mu^{-2}$ for both types of WSMs (assuming $\mu_s=0$), at variance with the Drude conductivity which is proportional to $\mu^2$. This implies that the nonlinear Hall effect becomes sizable when the Fermi energy approaches the energy of the Weyl nodes; such enhancement originates from the singularity of the Berry curvature at a Weyl node. The divergence of $\kappa^s$ as the Fermi energy falls on the Weyl nodes, however, can be evaded by the disorder-induced energy broadening, which imposes a lower bound on the Fermi energy $\mu \gtrsim \hbar/\tau$~\cite{morimoto16PRB_semi-cls-WSM}. Also, in our semiclassical treatment, we have neglected the interband transitions by restricting the external electric field to satisfy $eE\tau/\hbar <\mu/\hbar v_F$, leading to another constraint $\mu \gtrsim eE\tau v_F$. 
 
Furthermore, from a practical point of view, the strong dependence of the CNH effect on the Fermi energy suggests that when estimating the size of this effect in a real WSM material, one can just take into account those Weyl nodes in the Brillouin zone whose energies are very close to the Fermi energy for convenience. In this case, under a further assumption that these Weyl nodes are reasonably well separated, the Fermi surface is a set of disconnected regions about each Weyl node and hence the contributions to the CNH effect from each Weyl node just add up. Therefore, our formulation becomes more applicable under such consideration.

\subsection{\label{sec3B} Tilt dependence of the CNH response function}

 \begin{figure}[t]
    \centering
	\includegraphics[width=\linewidth]{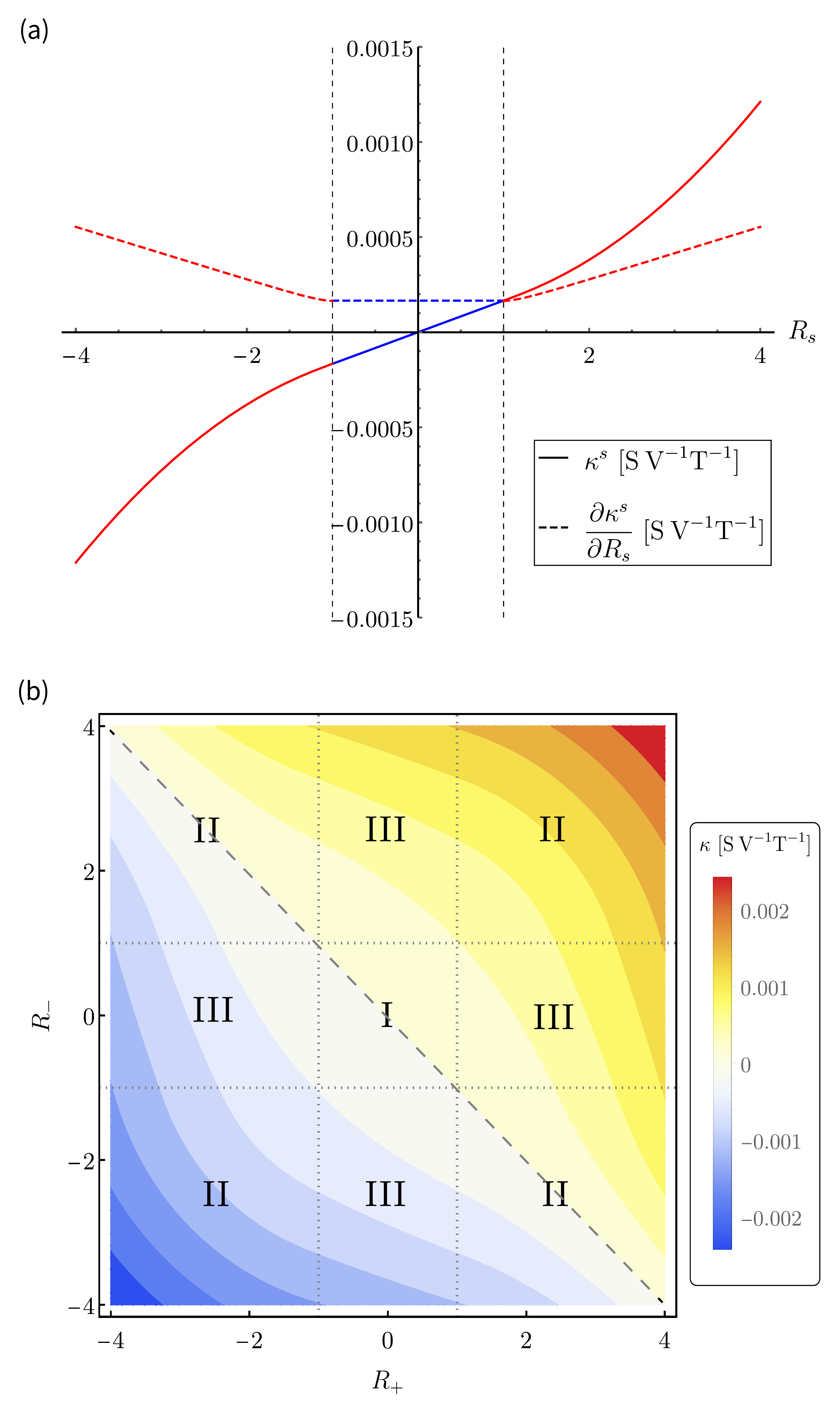}
    \caption{(a) The nonlinear current response function $\kappa^s$ (solid curve) and its first derivative with respect to the tilt parameter $R_s$ (dashed curve) as a function of $R_s$, with $\tau=10^{-13}$~s, $v_F=3\times 10^5$~m/s, $\mu-\mu_s = 10$~meV, and $\tilde\Lambda =10$. $\abs{R_s}<1$ (shown in blue) corresponds to type-I WSMs, whereas $\abs{R_s}>1$ (shown in red) corresponds to type-II WSMs. (b) Contour plot of the total response function for a pair of Weyl nodes, $\kappa = \kappa^+ + \kappa^-$, as a function of $R_+$ and $R_-$. Region I (II) in the parameter space corresponds to the case where both Weyl nodes are type-I (type-II). Region III represents the case where one of the Weyl nodes is type-I and the other is type-II. The dashed line corresponds to the case $R_+ = -R_-$.}
    \label{fig1}
\end{figure}

 To understand the tilt dependence of the nonlinear Hall response function, in Fig.~\hyperref[fig1]{2(a)} we show $\kappa^s$ for each Weyl node in a pair and its derivative with respect to the tilt parameter, $\pdv*{\kappa^s}{R_s}$, as a function of $R_s$, assuming $\mu-\mu_s=10$ meV (with $\mu_+=\mu_-$) and $\tilde\Lambda=10$. The phase transition from type-I (blue region) to type-II (red region) WSM can be clearly seen from the derivative of $\kappa^s$ with respect to $R_s$ (dashed curve) due to the discontinuity at $R_s=\pm 1$.  For both type-I and type-II WSMs, the individual contribution to the nonlinear Hall effect becomes more prominent as tilting of the Weyl cone gets larger due to the monotonic nature of the nonlinear current response function $\kappa^s$ (solid curve). Moreover, it is clear that $\kappa^s$ overall is an odd function of $R_s$, suggesting that the CNH effect vanishes when a pair of Weyl cones are oppositely tilted and their energies are the same, which is the case depicted in Fig.~\hyperref[schematic]{1(b)}.

In Fig.~\hyperref[fig1]{2(b)}, we consider a pair of Weyl nodes of opposite chiralities ($s=\pm 1$) and the total nonlinear current response  $\kappa = \kappa^+ +\kappa^-$ as a function of $R_+$ and $R_-$ is shown in the contour plot. Region I (II) in the parameter space corresponds to the case where both Weyl nodes are type-I (type-II). Region III corresponds to the case where one of the Weyl nodes is type-I and the other is type-II. When $R_+$ and $R_-$ have the same sign, the magnitude of $\kappa$ increases as the magnitude of either of the tilt parameters increases. On the other hand, when they have opposite signs, the magnitude of $\kappa$ first decreases as the magnitude of one of the tilt parameters increases while the other fixed. It then increases after the tilt parameter crosses the line $R_+=-R_-$. In particular, on the dashed line corresponding to $R_+ = -R_-$, $\kappa = 0$ in both type-I and -II WSMs due to a symmetric overall Fermi surface as shown in Fig.~\hyperref[schematic]{1(b)}, reaffirming that $\kappa^s$ is an odd function of $R_s$.

\subsection{\label{sec3C} Relative energy shift of Weyl nodes}

\begin{figure}[t]
    \centering
	\includegraphics[width=\linewidth]{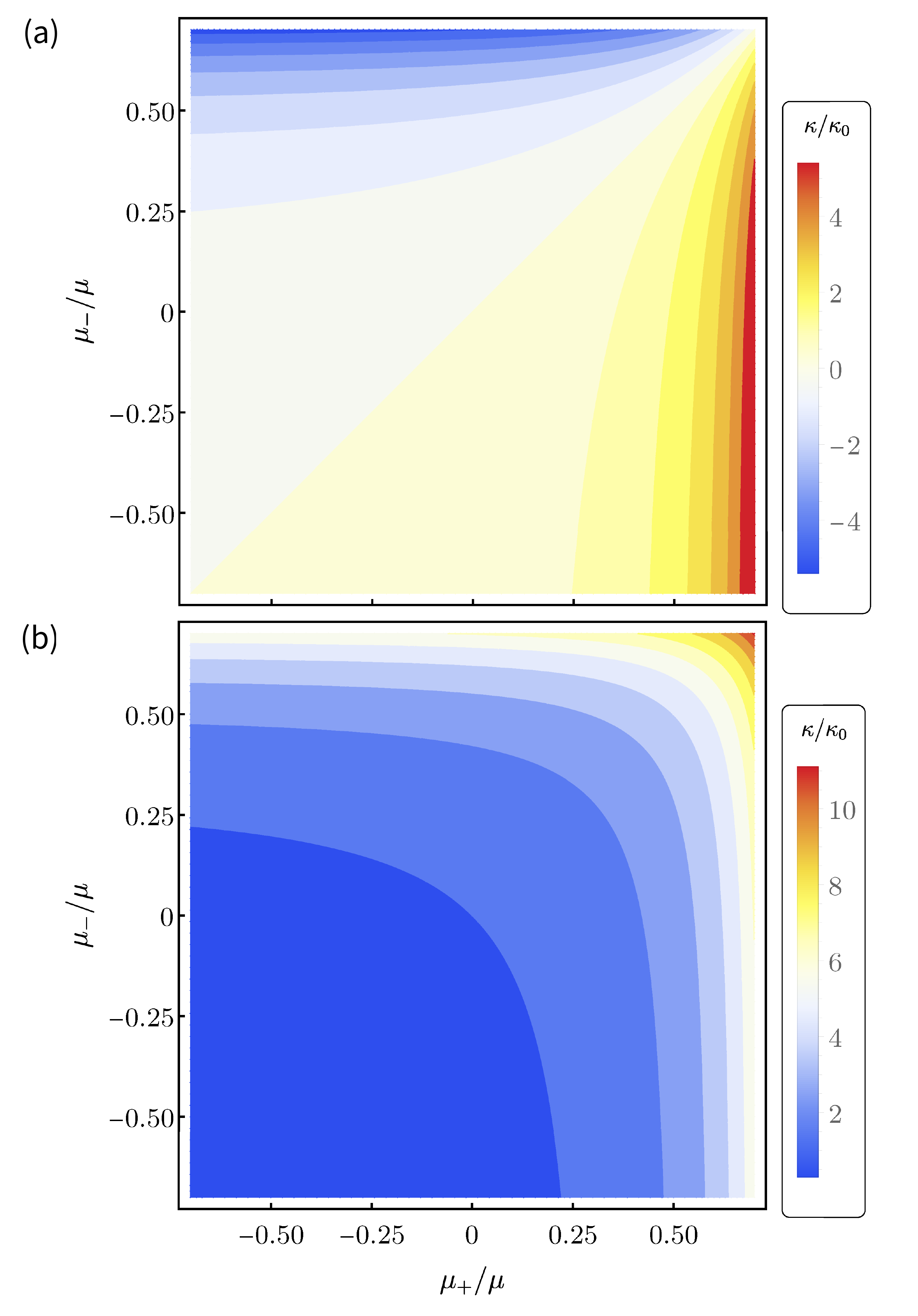}
    \caption{The ratio $\kappa/\kappa_0$ as a function of the energies $\mu_+$ and $\mu_-$ (normalized to the Fermi energy $\mu >0$) of a pair of Weyl nodes of opposite chiralities tilted along (a) opposite directions with $R_+=-R_-=0.5$ and (b) the same direction with $R_+=R_-=0.5$, where $\kappa = \kappa^+ + \kappa^-$ is the total contribution to the CNH from the pair of Weyl nodes and $\kappa_0\equiv \frac{3e^4v_F^2\tau}{40 \pi^2\hbar \mu^2}$ which has the same dimension as $\kappa$. }
    \label{fig:kappa-mu}
\end{figure}

For a WSM with broken inversion symmetry, individual pairs of Weyl nodes with opposite chiralities may be shifted to different energies \cite{Zyuzin_PRB12_WSM}. As long as the Weyl cones are tilted, such relative energy shift would modulate the CNH current due to the $(\mu-\mu_s)^{-2}$ dependence of the nonlinear response function $\kappa_s$ as given by Eqs.~(\ref{eq10}) and (\ref{eq14}).

For instance, in Fig.~\hyperref[fig:kappa-mu]{3(a)} we show the total nonlinear current response $\kappa \equiv \kappa^+ + \kappa^-$ as a function of the energy shifts $\mu_+$ and $\mu_-$ (normalized to a positive Fermi energy) of a pair of type-I Weyl nodes with opposite tilting. While $\kappa$ vanishes (as indicated by the diagonal line) when the Weyl nodes lie at the same energy whereby the two electron pockets are symmetric about the midpoint of the two nodes, a nonvanishing $\kappa$ arises when a relative energy shift between the pair of Weyl nodes is induced, which changes sign as the sign of $\mu_+-\mu_-$ is reversed. 

For a pair of Weyl nodes tilted along the same direction, as shown in Fig.~\hyperref[fig:kappa-mu]{3(b)}, the magnitude of $\kappa$ also varies with the relative energy shift between the Weyl nodes but the sign of $\kappa$ remains unchanged regardless of the sign of $\mu_+-\mu_-$, as opposed to the opposite tilting case.

\subsection{\label{sec3D} Comparison with other nonlinear Hall effects}

We are now in position to compare the CNH effect with other nonlinear Hall effects that were discovered previously. A nonlinear Hall conductivity linearly proportional to both electric and magnetic fields was derived by Morimoto and coworkers~\cite{morimoto16PRB_semi-cls-WSM} for Weyl fermions with linear and isotropic dispersion, which is finite when the electric and magnetic fields are perpendicular and hence does not emanate from the chiral anomaly. Another nonlinear Hall effect may be induced by the intrinsic Berry curvature dipoles~\cite{spivak09arxiv_Berry-dipole, lFu15PRL_nonlinearHall, Low15PRB_NLH, Du18PRL_NLH_WTe2, Ma19nature_NL-Hall, jShan19NatMater_NL-Hall} or disorder \cite{Du2019NatComm_disorderNLH,xcXie20arxiv_NLH} in time-reversal-invariant systems. The corresponding Berry-curvature-dipole-induced nonlinear Hall (BNH) current density can be expressed as $j_{a}^\text{BNH}=\sum_s\chi^s_{abc}E_{b}E_{c}$ with the nonlinear response function~\cite{spivak09arxiv_Berry-dipole,lFu15PRL_nonlinearHall}
\begin{equation} 
\chi^s_{abc}=\epsilon_{acd}\frac{e^{3}\tau }{2}\int_{\vb k}f^{s}_{0}\frac{\partial }{\partial k_{b}}\Omega^s_{d}
\label{chi-BNH}
\end{equation}
where the index $s$ represents the chirality of the Weyl nodes and the rest of the indices represent spatial components, $\epsilon_{acd}$ is the three-dimensional Levi-Civita antisymmetric tensor, and the Berry curvature dipole is defined by the integral. Note that a magnetic field is not necessary for a nonvanishing Berry curvature dipole; in other words, the corresponding nonlinear Hall effect in WSMs does not rely on the chiral anomaly. For comparison, it is also instructive to rewrite the CNH current given by Eq.~\eqref{eq7} in component form as $j_{a}^\text{CNH} = \sum_s\varkappa^s_{abcd}E_{b} E_{c} B_{d}$, with
\be
\varkappa^s_{abcd}=\epsilon _{abl} \epsilon _{gcm} \epsilon_{gdn}  \f{e^4\tau}{\hbar^3}\int_{\vb k}f^{s}_{0}\frac{\partial }{\partial
k_{n}}\left( \Omega^s_{l}\Omega^s_{m}\right).
\label{Eq:T}
\ee
Comparing Eq.~(\ref{chi-BNH}) with (\ref{Eq:T}), it is evident that the CNH response function $\varkappa^s_{abcd}$ is intrinsically different from the response function due to the Berry curvature dipole. 

In the presence of an external magnetic field, however, the orbital magnetic moment leads to a correction to the equilibrium distribution function; that is, $f_0^s(\tilde{\varepsilon}^s_{\vb k})\simeq f_0^s(\varepsilon^s_{\vb k})-\pdv{f^s_0}{\varepsilon^s_{\vb k}} \qty(\vb{m}^s_{\vb k} \vdot \vb B)$ in the small-field limit. When the second term enters Eq.~\eqref{chi-BNH}, it gives rise to a contribution to the nonlinear Hall current density that has the same form as Eq.~\eqref{eq:J^CNH}. The difference is that the nonlinear response function due to the correction to the Berry curvature dipole comes with an opposite sign compared with the CNH response function and its magnitude is about 1/3 of the CNH counterpart. Consequently, the BNH contribution will not interfere with the detection of the CNH effect as the net effect is still CNH-dominant, despite the coexistence of the two when a magnetic field is applied.


\subsection{\label{sec3E} Materials and experimental considerations}

To discuss the experimental prospects of the CNH effect, we first give an order-of-magnitude estimate of the size of this effect for a pair of Weyl cones, which are assumed to tilt in the same direction. For reference, we compare the size of the nonlinear Hall conductivity, $\sigma_\text{I(II)} = \kappa_\text{I(II)}\vb E\vdot\vb B$, to the size of the Drude conductivity $\sigma_D$ due to this pair of Weyl nodes, which is given in Ref.~\cite{He&Zhang19PRL_NPHE}. For a type-I WSM such as TaAs~\cite{hmWeng15PRX_WSM-noncen,Arnold16PRL}, using typical parameters $v_F = 3\times 10^5$ m/s, $\mu = 20$ meV, and assuming a tilt parameter $R_s = 0.1$ leads to a ratio of the nonlinear Hall conductivity to the Drude conductivity, $\sigma_\text{I}/\sigma_D \simeq 1\%$, for an electric field $E = 100$ V/cm applied in the $x$ direction and a parallel magnetic field $B = 9$ T. Similarly, a type-II WSM such as MoTe$_2$ \cite{Wang16PRLMoTe2, Huang16NatureMat} with $R_s = 1.5$ would lead to a ratio of the same order of magnitude. Tilting of Weyl cones in principle can be varied by applying strains \cite{Sun2015_MoTe2, Ruan2016_HgTe_WSM, Arjona2018_rot_strain}---an experimentally accessible knob to control the size of the CNH effect and to differentiate it from other linear and nonlinear Hall effects.

It should be noted that the order-of-magnitude estimate above assumes same tilting all the Weyl cones in a WSM for simplicity. To make a more accurate quantitative prediction of the CNH effect for a candidate WSM with broken inversion symmetry, one needs to take into account all the relevant pairs of Weyl nodes in the Brillouin zone; that is, those whose energies are close to the Fermi energy. For example, in the type-II WSM material LaAlGe \cite{Xu2017laalge}, there are totally 40 Weyl nodes in the bulk Brillouin zone, but only the 16 type-II Weyl nodes labeled as W$_2$ are relevant because the other type-I Weyl nodes are located further above the Fermi level and their tiltings are much smaller. The dispersion of each of the Weyl cones can in principle be obtained from \ti{ab initio} calculations and/or angle-resolved photoemission spectroscopy measurements in experiment, which can be used to fit the linear two-band model \eqref{eq2}. One can then apply the main results in this paper, Eqs.~(\ref{eq:J^CNH})--(\ref{eq14}), to evaluate the total CNH conductivity in a candidate material.

Experimentally, the CNH effect can be separated from linear Hall effects~\cite{aBurkov17PRB_PHE-WSM, sTewari17PRL_PHE-WSM,pLi&xZhang18PRB_PHE-WSM,Agarwal19PRB_tiltWSM} in WSMs: In ac measurements, this can be easily achieved by measuring the second-harmonic Hall resistance~\cite{Pan&Zhang18NatPhys_BMER,He&Zhang19PRL_NPHE} wherein linear Hall contributions are automatically excluded. In dc measurements, they can also be distinguished by proper alignment of the external electric and magnetic fields.

\section{\label{sec4}Conclusion and Outlook}

In this work, we proposed a nonlinear Hall effect in WSMs that arises from the combined actions of the chiral anomaly and the anomalous velocity, which we denoted by the CNH effect. We showed that this effect requires inversion-symmetry breaking as well as asymmetric Fermi surfaces, which can be achieved via tilting of the Weyl cones. It is worth mentioning that even though for simplicity, much of our analysis revolves around the simplest configuration---a pair of Weyl nodes of opposite chiralities tilted along the same axis, our results for the CNH current density can be generalized to any Weyl configuration with an arbitrary number of Weyl nodes that are tilted in arbitrary directions, provided that the Weyl nodes are well separated.

In addition, an asymmetric Fermi surface may be achieved by breaking certain symmetry (such as time reversal) in addition to inversion. We thus anticipate that the CNH effect can be observed in the family of noncentrosymmetric magnetic WSMs~\cite{Hasan18PRB_WSM-IB&TRB, Sanchez20_MNC_WSM}. The search for candidate WSM materials that may host sizable CNH effect would be desirable to be pursued in future investigations with inputs of \ti{ab initio} calculations.

\section*{Acknowledgment}
S. S.-L. Z. is grateful to T.-R. Chang, G. Bian, and L.-L. Wang for helpful discussions. Work by R.-H. L. and S. S.-L. Z. were supported by College of Arts and Sciences, Case Western Reserve University. A. A. B., and O. G. H. were supported by Center for Advancement of Topological Semimetals,
an Energy Frontier Research Center funded by the U.S.
Department of Energy Office of Science, Office of Basic
Energy Sciences, through the Ames Laboratory under
Contract No. DE-AC02-07CH11358.

\appendix

\section{\label{appx-A} Four-band Weyl Hamiltonian and Asymmetric Fermi Surfaces}

\begin{figure*}[!htp]
    \centering

	\includegraphics[width=\linewidth]{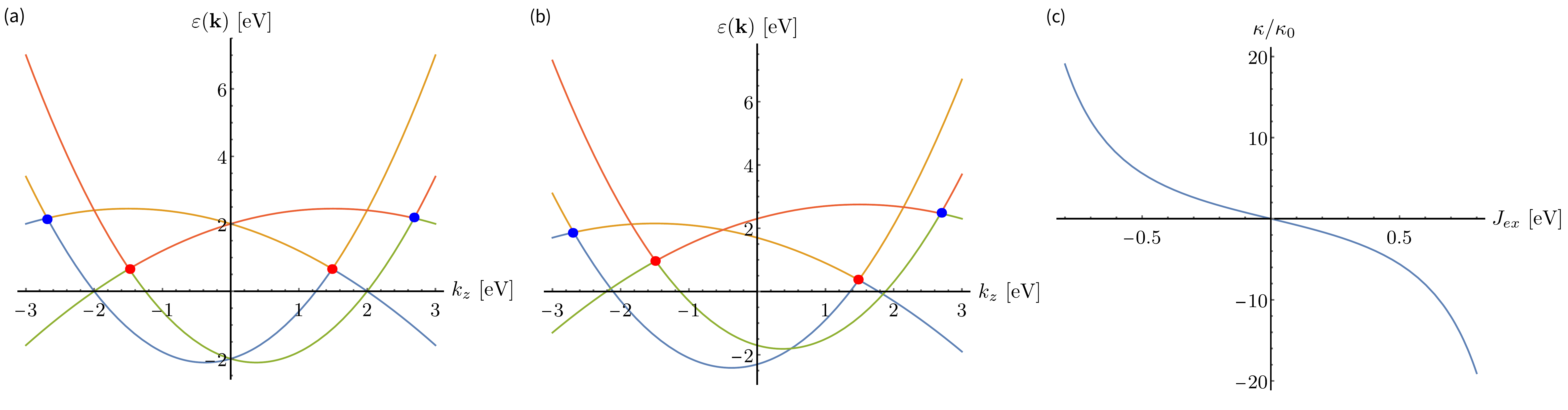}
    \caption{(a) \& (b) Schematics of the energy dispersion corresponding to the four-band Hamiltonian \eqref{eq-A:full-H} with $k_x=k_y=0$. For illustrative purposes, we take $v_F = 1$, $\lambda=0.5\ \text{eV}^{-1}$, and $Q_D=2\ \text{eV}$. (a) When the inversion-breaking parameter $v_I = 0.6$ but $J_{ex} = 0$, each of the Dirac points splits into two Weyl points, generating two pairs of Weyl nodes on the $k_z$-axis. Along with $R= 0.3\ \text{eV}^{-1}$, the energies of the Weyl nodes are lifted and the surrounding dispersion becomes tilted. (b). When the time-reversal-breaking exchange coupling $J_{ex} = 0.3\ \text{eV}$ along with $v_I= 0.6$, the energies of Weyl nodes of the same chirality are displaced by opposite amounts, leading to an asymmetric overall Fermi surface. The blue and red dots denote the chirality ($\pm 1$) of the Weyl nodes. (c) The ratio $\kappa/\kappa_0$ as a function of the exchange coupling $J_{ex}$, where $\kappa$ is the total nonlinear Hall response function for the four Weyl nodes calculated based on Eq.~\eqref{eq10} in the main text and $\kappa_0\equiv \frac{3e^4v_F^2\tau}{40 \pi^2\hbar \mu^2}$ with $\mu = 2$ eV.}
    \label{figd1}
\end{figure*}

As illustrated in Fig.~\ref{schematic} in the main text, a finite CNH effect can be hosted in WSMs with broken inversion symmetry when tilting of Weyl cones leads to a Fermi surface that is asymmetric about the $\Gamma$ point. The desired configuration of tilting may be achieved by breaking additional symmetries such as the time reversal, as we will demonstrate below. 

Let us commence with a more general low-energy model Hamiltonian for a Dirac semimetal (e.g., Cd$_3$As$_2$) with two Dirac points at $(0,0,\pm Q_D)$:
\be\label{eq-A:DSM}
H_0(\mathbf{k}) =R k_z^2+ v_F\left[\sigma_x s_z k_x -\sigma_y k_y-\lambda\left(k_z^2-Q_D^2\right)\sigma_z\right]
\ee
where the term $R k_z^2$ breaks the particle-hole symmetry, leading to tilted energy dispersion around the Dirac points and shifting their energy~\footnote{Note that since the particle-hole asymmetric term is given by $R k_z^2$ times a $4\times 4$ identity matrix which does not affect the eigenvectors, it does not modify the Berry curvature, nor does the resulting tilting term.},  $\sigma_i$ and $s_i$ ($i=x,y,z$) are the Pauli matrices for the orbital and spin degrees of freedom, respectively, the material parameters $v_F$, $\lambda$ and $R$ are all independent of $\vb{k}$, and we have adopted the natural units $\hbar=c=1$ for the ease of notation. 

One can break the inversion symmetry $\mathcal P$ by adding a term
\be \label{eq-A:IB}
H_\text{IB}=v_I\sigma_z s_z k_z
\ee
to the Dirac Hamiltonian (\ref{eq-A:DSM}) ($\mathcal P^{-1}H_\text{IB}(\vb k) \mathcal P\neq H_\text{IB}(-\vb k)$ with $\mathcal P = \sigma_z$). By doing so, each of the two Dirac points split into two Weyl points with opposite chiralities that also lie on the $z$ axis, as shown in Fig.~\hyperref[figd1]{4(a)}.~\footnote{Note that the additional band crossings in the spectrum away from the $\Gamma$ point are nodal lines, which can be gapped out by breaking the rotational symmetry of Cd$_3$As$_2$ with a term such as $m k_z\sigma_x s_x$. In what follows, we will ignore these nodal lines and focus on the physics related to the Weyl nodes.} Note that as the overall Fermi surface turns out to be symmetric about the $\Gamma$ point, we would expect the CNH effect to vanish according to our analysis in the main text.

One simply way to obtain an asymmetric Fermi surface is by adding another term 
\be
H_\text{TRB}=J_{ex}s_z
\ee
to further break the time-reversal symmetry ($\mathcal T^{-1}H_\text{TRB}(\vb k) \mathcal T\neq H_\text{TRB}(-\vb k)$ with $\mathcal T = is_y \mathcal K$, where $\mathcal K$ is the complex conjugation operator). The full Weyl Hamiltonian then becomes
\be\label{eq-A:full-H}
H = H_0+H_\text{IB}+H_\text{TRB}\,,
\ee
The term $H_\text{TRB}$ may arise from the exchange interaction between the Weyl-fermion spin and the (uniform) magnetization in a magnetic WSM with $J_{ex}$ the exchange coupling coefficient. It leads to a relative energy shift between Weyl nodes of the same chirality, resulting in an asymmetric Fermi surface, as shown in Fig.~\hyperref[figd1]{4(b)}. We thus anticipate the emergence of a  nonvanishing CNH effect. In what follows, we explicitly calculate the total CNH response function $\kappa$ by linearizing the four-band Hamiltonian (\ref{eq-A:full-H}) and applying Eqs.~(\ref{eq10}) and (\ref{eq14}) in the main text.

The band dispersion corresponding to the full Hamiltonian $H=H_0+H_\text{IB}+H_\text{TRB}$ is
\begin{eqnarray}\label{eqd2}
\varepsilon_{\zeta,\nu} (\vb k) = &&\ \zeta v_F\sqrt{k_x^2+k_y^2+\qty[\tilde v k_z+\nu \lambda \left(k_z^2-Q_D^2\right)]^2}\nonumber\\ 
 &&\ +Rk_z^2-\nu J_{ex},
\end{eqnarray}
where $\tilde v\equiv v_I/v_F$ and $\zeta,\nu=\pm 1$. It describes two pairs of Weyl nodes---one situated at $(0,0, (Q_D^2+q_I^2)^{1/2} \pm q_I)$ and the other at $(0,0, -(Q_D^2+q_I^2)^{1/2}\pm q_I)$, where $q_I\equiv \tilde v/2\lambda = v_I/2\lambda v_F$. To make a connection with the linearized one-node Hamiltonian \eqref{eq1} in the main text, we may do an expansion around a pair of Weyl nodes located at, say, $(0,0,  s q_I+(Q_D^2+q_I^2)^{1/2})$, where $s=\pm 1$ denotes the chirality. Upon closer inspection, we find that the Weyl node with $s=1$ is the touching point between the bands $\varepsilon_{-1,-1}$ and $\varepsilon_{1,-1}$ ($\nu=-1$), while the one with $s=-1$ is the touching point between $\varepsilon_{-1,1}$ and $\varepsilon_{1,1}$ ($\nu= 1$). So we establish that $\nu = -s$ for this pair of Weyl nodes. By expanding the momenta around the Weyl nodes, $(k_x,k_y,k_z)\simeq (\tilde k_x, \tilde k_y, sq_I+(q_I^2+Q_D^2)^{1/2}+\tilde k_z)$, and substituting it into Eq.~\eqref{eqd2}, we arrive at the energy dispersion in the vicinity of the Weyl node of chirality $s$,
\begin{eqnarray}\label{eqd3}
\hspace{-2em}
	\varepsilon^s_{\tilde{\vb k},1} \simeq &&\ \pm v_F\sqrt{\tilde k_x^2+\tilde k_y^2+\lambda^2\qty(2\sqrt{q_I^2+Q_D^2}\tilde k_z+\tilde k_z^2)^2}\nonumber\\
	&&\ +R\qty(2q_I^2+Q_D^2+2sq_I\sqrt{q_I^2+Q_D^2})+ s J_{ex} \nonumber\\
	&&\ +2R\qty(sq_I+\sqrt{q_I^2+Q_D^2})\tilde k_z+R\tilde k_z^2 \,.
\end{eqnarray}
If one keeps only terms up to the first order in $\tilde k_i$, the dispersion above matches the dispersion \eqref{eq2} given by Hamiltonian \eqref{eq1} describing the low-energy physics around the Weyl point situated at $(0,0,sq_I+(q_I^2+Q_D^2)^{1/2})$, provided that $\tilde v^2+4\lambda^2 Q_D^2=1$, $R_s = \f{R}{\lambda v_F}(s\tilde v+1)$, and $\mu_s=\f{R}{4\lambda^2}(\tilde v+s)^2-sJ_{ex}$, where $\tilde v\equiv v_I/v_F$ and $\hbar =1$ is assumed. 

Similarly, for the other pair of Weyl nodes located at $(0,0,-sq_I-(q_I^2+Q_D^2)^{1/2})$, we identify $\nu = s$ in the dispersion. Doing the same expansion around the Weyl nodes leads to the following low-energy dispersion,
\begin{eqnarray}\label{eqd4}
\hspace{-1.5em}
	\varepsilon^s_{\tilde{\vb k},2} \simeq &&\ \pm v_F\sqrt{\tilde k_x^2+\tilde k_y^2+\lambda^2\qty(\tilde k_z^2-2\sqrt{q_I^2+Q_D^2}\tilde k_z)^2}\nonumber\\
	&&\ +R\qty(2q_I^2+Q_D^2+2sq_I\sqrt{q_I^2+Q_D^2})- sJ_{ex}\nonumber\\
	&&\ -2R\qty(sq_I+\sqrt{q_I^2+Q_D^2})\tilde k_z +R\tilde k_z^2\,.
\end{eqnarray}
Matching the linearized dispersion \eqref{eq2}, we find $\tilde v^2+4\lambda^2 Q_D^2=1$, $R_s = -\f{R}{\lambda v_F}(s\tilde v+1)$, and $\mu_s=\f{R}{4\lambda^2}(\tilde v+s)^2+sJ_{ex}$ for this pair of Weyl nodes. 

Placing the parameters $R_s$ and $\mu_s$ derived above for each of the four Weyl nodes into Eqs.~(\ref{eq10}) and (\ref{eq14}) in the main text, we calculate the contribution from individual Weyl nodes to the CNH conductivity and then sum them up to obtain the total CNH conductivity. In Fig.~\hyperref[figd1]{4(c)}, we show the dependence of the total CNH response function $\kappa$ on the exchange coupling strength $J_{ex}$, which confirms that a finite $\kappa$ appears when $J_{ex}\neq 0$ whereby the Fermi surface is asymmetric about $\vb k=0$.  


\section{\label{appx-B} Dependence of the Nonlinear Response Functions on the Separation between Weyl Nodes}
Here we consider two Weyl nodes of opposite chiralities lying on the $z$ axis, separated by $2Q$ in momentum space. Again, the Weyl cones are assumed to be tilted along the $z$ axis. The low-energy Hamiltonian for a single Weyl node that takes into account the separation is given by
\be \label{eqb1}
H^s(\vb k) = \hbar v_F\qty[s(\vb k-sQ\vu e_z)\vdot \boldsymbol\sigma + R_s (k_z-sQ)\sigma_0]+\mu_s.
\ee
The Berry curvature corresponding to this Hamiltonian is
\be
\boldsymbol{\Omega}^{s}(\vb k) = -s\frac{\pm(\vb k-sQ\vu e_z)}{2\abs{\vb k-sQ\vu e_z}^3}.
\ee
With Eq.~\eqref{eq6} in the main text and the fact that a nonlinear response to an external electric field can only arise from terms in that involve $f_1^s$ and $f_2^s$ in the expansion of the distribution function, the nonlinear current-density components can be written in the cylindrical coordinates as
\begin{widetext}
\be \label{eqs11}
j_i = \sum_s \int_0^\infty k_\perp dk_\perp \int_0^{2\pi}d\phi\int_{-\Lambda_z-sQ}^{\Lambda_z-sQ}d\tilde k_z^s\:\mathcal M^s_i(k_\perp,\phi,\tilde k_{z}^s)\:\delta\qty[\mu-\hbar v_F\qty(R_s\tilde k_z^s\pm\sqrt{\big(k_\perp\big)^2+\big(\tilde k_z^s\big)^2})-\mu_s],
\ee
\end{widetext}
where $k_\perp\equiv (k_x^2+k_y^2)^{1/2}$, $\tilde k_z^s\equiv k_z-sQ$, and $\mathcal M^{s}_i(k_\perp,\phi,\tilde k_z^s)$ is a function proportional to the squares of the electric-field strength that comes from expanding the integrand in Eq.~\eqref{eq6}. 

As mentioned in the main text, a momentum cutoff must be introduced for calculations of type-II WSM contributions within the linear model. With the cylindrical geometry, the momentum cutoff $\Lambda_z$ for a type-II WSM is imposed on $k_z$. Similar to what is discussed in Ref.~\cite{Agarwal19PRB_tiltWSM}, analyzing the roots of the Dirac $\delta$ function helps determine the integration limits for type-I and type-II WSMs. In this case, the root for the conduction band can be written as
\be
k_\perp = \sqrt{\qty[(R_s-1)\tilde k_{z}^s-\f{\mu-\mu_s}{\hbar v_F}]\qty[(R_s+1)\tilde k_{z}^s-\f{\mu-\mu_s}{\hbar v_F}]},
\ee
and for the valence band, 
\be
k_\perp = \sqrt{\qty[\f{\mu-\mu_s}{\hbar v_F}+(1-R_s)\tilde k_{z}^s]\qty[\f{\mu-\mu_s}{\hbar v_F}-(1+R_s)\tilde k_{z}^s]}.
\ee
Therefore, assuming $\mu>\mu_s$, for type-I WSMs ($\abs{R_s}<1$), the requirement that $k_\perp$ is real gives the following limits for $\tilde k_z^s$:
\be
-\f{\mu-\mu_s}{\hbar v_F(1-R_s)}\leqslant \tilde k_z^s \leqslant \f{\mu-\mu_s}{\hbar v_F(1+R_s)}.
\ee

For type-II WSMs with $R_s>1$, with the cutoff imposed, the limits for the conduction band are
\be
-(\Lambda_z+sQ)\leqslant \tilde k_z^s\leqslant \f{\mu-\mu_s}{\hbar v_F(R_s+1)},
\ee 
whereas those for the valence band are
\be
\f{\mu-\mu_s}{\hbar v_F(R_s-1)}\leqslant \tilde k_z^s\leqslant \Lambda_z-sQ.
\ee 
Similarly, for type-II WSMs with $R_s<-1$, we obtain
\be
\f{\mu-\mu_s}{\hbar v_F(R_s-1)}\leqslant \tilde k_z^s\leqslant \Lambda_z-sQ
\ee
for the conduction band and
\be
-(\Lambda_z+sQ)\leqslant \tilde k_z^s \leqslant \f{\mu-\mu_s}{\hbar v_F(R_s+1)}
\ee
for the valence band. The case of $\mu<\mu_s$ can be worked out similarly. 

Using the integration limits above in Eq.~\eqref{eqs11}, we find that the nonlinear response function for type-I WSMs is given by
\be
\kappa_\text{I}^s = \f{3e^4v_F^2\tau(\mu-\mu_s)}{40\pi^2\hbar\abs{\mu-\mu_s}^3}R_s,
\ee
the same as Eq.~\eqref{eq10} in the main text. For type-II WSMs, it is given by
\begin{eqnarray}\label{eqB4}
	\kappa^s_\text{II} =\ &&\f{5\kappa^s_\text{I}}{12\abs{R_s}^5}\Bigg[\qty(R_s^6+\f{3}{2}R_s^4-\f{1}{10})\nonumber\\&&-\f{R_s^2-1}{2}\qty(\delta_+^2+\delta_-^2)+(R_s^2-1)\qty(\delta_+^3+\delta_-^3)\nonumber\\
	&&-\f{2R_s^2-3}{4}\qty(\delta_+^4+\delta_-^4)-\f{1}{5}\qty(\delta_+^5+\delta_-^5)\Bigg],
\end{eqnarray}
where
\be
	\delta_{\pm} = \qty[1+R_s(s\tilde Q\pm\tilde\Lambda_z)]^{-1},
\ee
with $\tilde Q\equiv \hbar v_F Q/(\mu-\mu_s)$ and $\tilde\Lambda_z\equiv \hbar v_F \Lambda_z/(\mu-\mu_s)$. 

It is worth noting that the cutoff-independent terms in Eq.~\eqref{eqB4} agree with those in Eq.~\eqref{eq14} in the main text, and they dominate when $\tilde\Lambda_z\gg 1$. In this case, the result is independent of both $Q$ and $\Lambda_z$. The $Q$-dependence of the nonlinear response function is fairly weak except for values of $Q$ close to the momentum cutoff $\tilde\Lambda_z$. For brevity, we adopt a simpler Hamiltonian without $Q$ dependence for our calculations presented in the main text.


\section{\label{appx-C} Contributions of Intra- and Inter-node Scatterings to the CNH Effect}

With the relaxation-time approximation, the scattering term on the right-hand side of the Boltzmann equation~\eqref{eq7new} in the main text is characterized by the intranode scattering time $\tau$. To include the effect of internode scattering between two Weyl nodes of opposite chiralities ($s=\pm 1$), we can write down the following set of coupled Boltzmann equations,
\be
\dot{\vb k}^s\vdot\pdv{f^s}{\vb k}  = -\f{f^s - \expval{f^s}}{\tau} - \f{\expval{f^s}-\expval{f^{-s}}}{\tau_\text{inter}},
\ee
where $\tau,\tau_\text{inter}$ denote intranode and internode scattering times, respectively, and $\expval{f} \equiv \int\dd\Omega_{\vb k}\: f/4\pi$ denotes the angular average of the distribution $f$ over $\vb k$ space.

To obtain nonequilibrium distribution at the first order in $E$, $f^s_1$, we start with the Boltzmann equation for Weyl fermions with positive chirality.
With $\pdv{f^+_0}{\vb k} = \pdv{\varepsilon^+}{\vb k}\pdv{f^+_0}{\varepsilon^+} = -\hbar\vb v^+ \delta(\mu-\varepsilon^+)$ and Eq.~\eqref{eq4b} in the main text, we find
\begin{eqnarray}\label{s29}
f^+_1 &&\simeq \expval{f^+_1} -\tau \delta(\mu-\varepsilon^+)\vb v^+\vdot\Bigg[e\vb E +\f{e^2}{\hbar}(\vb E\vdot\vb B)\vb\Omega^+\nonumber\\&&- \f{e^2}{\hbar}(\vb B\vdot\vb\Omega^+)\vb E\Bigg]-\f{\tau}{\tau_\text{inter}}\qty(\expval{f^+_1}-\expval{f^-_1}).
\end{eqnarray}
For an untilted WSM, substituting Eqs.~\eqref{eq2new} and \eqref{eq2} in the main text into Eq.~\eqref{s29} and taking the angular average on both sides gives
\be
\expval{f^+_1}-\expval{f^-_1} = \f{\tau_\text{inter} e^2 v_F}{3\hbar k^2}(\vb E\vdot\vb B)\delta(\mu-\varepsilon^+).
\ee
The above result suggests the fact that the imbalance of the electron distribution between the two Weyl nodes is a consequence of the interplay between the chiral anomaly represented by $\vb E\vdot\vb B$ and the internode scattering characterized by $\tau_\text{inter}$. Performing the same procedure on the Boltzmann equation for $f^-$ leads to a similar result. To solve for $\expval{f^+_1}$ and $\expval{f^-_1}$ individually, one will need another condition, that is, conservation of the total electron density. However, as will be seen below, to determine the effect of the internode scattering on the CNH effect in tilted WSMs, the explicit expressions of them are not necessary.

With Eq.~\eqref{eq6a} in the main text, the CNH current density for the positive node can be written as
\be
\vb j^+ = -e\int\f{\dd^3k}{(2\pi)^3}\qty(\f{e}{\hbar}\vb E\cp\vb\Omega^+) f^+_1.
\ee
Since $\boldsymbol{\Omega}^+$ is odd and all $\expval{f}$'s are even under $\vb k\to -\vb k$, it is obvious that only the second term within square brackets on the right-hand side of Eq.~\eqref{s29} will contribute to a finite current density in tilted WSMs, and is proportional to the intranode scattering time. On the other hand, the terms related to internode scattering do not contribute. The same argument applies to $\vb j^-$. Hence, the above analysis justifies the neglect of internode scattering in our formalism.\\

\section{\label{appx-D} Calculation of the Nonlinear Current Density with Spherical Geometry}

Similar to Eq.~\eqref{eqs11}, exploiting the spherical symmetry of the system described by Hamiltonian~\eqref{eq1} in the main text, the nonlinear current-density components can be written as
\begin{widetext}
\be \label{eq8}
j_{i} = \sum_s \int_0^\infty k^2  dk\int_0^{2\pi}d\phi \int_{-1}^{1}d (\cos\theta)\:\mathcal M^{s}_i(k, \phi,\cos\theta)\:\delta\qty[\mu-\hbar v_F k\qty(R_s\cos\theta\pm 1)-\mu_s],
\ee
\end{widetext}
where $k = \abs{\vb k}$. In this case, the root for the conduction band in the Dirac $\delta$ function is given by
\be
k = \f{\mu-\mu_s}{\hbar v_F(R_s\cos\theta+1)},
\ee
whereas for the valence band,
\be
k = \f{\mu-\mu_s}{\hbar v_F(R_s\cos\theta-1)}.
\ee
Since $k\geq 0$ and $\mu>\mu_s$ is assumed, we must have $R_s\cos\theta+1 >0$ for the conduction band and $R_s\cos\theta-1 >0$ for the valence band. Then for type-I WSMs with $\abs{R_s}<1$, the integration range of $\cos\theta$ is from $-1$ to 1 (only the conduction band is taken into account for $\mu >\mu_s$). Similarly, the same integration range holds for the valence band when $\mu <\mu_s$.

The situation is different for type-II WSMs ($\abs{R_s}>1$). As mentioned in the main text, to do sensible calculations for type-II WSMs one needs to impose a radial momentum cutoff $\Lambda$ such that $k\leqslant\Lambda$. This cutoff then translates into a change in the integration limits of $\cos\theta$. For $R_s > 1$, requiring $k\leqslant\Lambda$ leads to the following integration limits for the conduction band:
\be
\qty(\f{\mu-\mu_s}{\hbar v_F\Lambda}-1)\f{1}{R_s}\leqslant \cos\theta \leqslant 1,
\ee
while for the valence band the limits are
\be
\qty(\f{\mu-\mu_s}{\hbar v_F\Lambda}+1)\f{1}{R_s}\leqslant \cos\theta \leqslant 1.
\ee
On the other hand, for $R_s<-1$, we get the following limits for the conduction band:
\be
-1\leqslant \cos\theta \leqslant \qty(\f{\mu-\mu_s}{\hbar v_F\Lambda}-1)\f{1}{R_s},
\ee
and for the valence band:
\be
-1\leqslant \cos\theta \leqslant \qty(\f{\mu-\mu_s}{\hbar v_F\Lambda}+1)\f{1}{R_s}.
\ee
Using these limits along with Eq.~\eqref{eq8}, we calculated the CNH current density and hence the response functions reported in the main text.

\bibliographystyle{apsrev4-2}
\bibliography{CNH_refs}
\medskip

\end{document}